\pdfoutput=1

\documentclass[11pt]{article}

\usepackage{acl}

\usepackage{times}
\usepackage{latexsym}

\usepackage[T1]{fontenc}

\usepackage[utf8]{inputenc}

\usepackage{microtype}

\usepackage{inconsolata}
\usepackage{relsize} 
\usepackage{graphicx}
\usepackage{listings}
\usepackage{todonotes}
\usepackage{color}
\definecolor{Code}{rgb}{0,0,0}
\definecolor{Keywords}{rgb}{0.13,0.13,1}     
\definecolor{Strings}{rgb}{0.9,0.17,0.31}    
\definecolor{Comments}{rgb}{0.25,0.5,0.37}   
\definecolor{Numbers}{rgb}{0.17,0.57,0.68}   
\newcommand{\QueryTuner}{QueryExplorer}

\title{\QueryTuner{}: An Interactive Query Generation Assistant for Search and Exploration}

\author{Kaustubh D. Dhole\\
\textsmaller[1]{Department of Computer Science} \\
\textsmaller[1]{Emory University, Atlanta} \\
\textsmaller[1]{\texttt{kaustubh.dhole@emory.edu}} \\
  \And
   Shivam Bajaj\\
   \textsmaller[1]{Cox Communications} \\
\textsmaller[1]{Atlanta} \\
  \textsmaller[1]{\texttt{shivam.bajaj@cox.com}}\\ 
  \And
   Ramraj Chandradevan \hspace{50pt}\\
   \textsmaller[1]{Department of Computer Science} \hspace{50pt}\\
\textsmaller[1]{Emory University, Atlanta} \\
\textsmaller[1]{\texttt{rchan31@emory.edu}} 
  \\\And
  Eugene Agichtein\\
  \textsmaller[1]{Department of Computer Science}\hspace{-50pt}\\
\textsmaller[1]{Emory University, Atlanta} \\
  \textsmaller[1]{\texttt{eugene.agichtein@emory.edu}}
  \\ \hspace{-500pt}\\
  }

\begin{document}
\maketitle
\begin{abstract}
Formulating effective search queries remains a challenging task, particularly when users lack expertise in a specific domain or are not proficient in the language of the content. Providing example documents of interest might be easier for a user. However, such query-by-example scenarios are prone to concept drift, and the retrieval effectiveness is highly sensitive to the query generation method, without a clear way to incorporate user feedback. To enable exploration and to support Human-In-The-Loop experiments we propose \href{https://github.com/emory-irlab/query-explorer}{\textbf{\QueryTuner{}}}\footnote{\href{https://github.com/emory-irlab/query-explorer}{https://github.com/emory-irlab/query-explorer}} -- an interactive query generation, reformulation, and retrieval interface with support for HuggingFace generation models and PyTerrier's retrieval pipelines and datasets, and extensive logging of human feedback. To allow users to create and modify effective queries, our demo\footnote{\href{https://www.youtube.com/watch?v=sXBU8-uWR3o}{Demonstration Video of QueryExplorer}} supports complementary approaches of using LLMs interactively, assisting the user with edits and feedback at multiple stages of the query formulation process. With support for recording fine-grained interactions and user annotations, \QueryTuner{} can serve as a valuable experimental and research platform for annotation, qualitative evaluation, and conducting Human-in-the-Loop (HITL) experiments for complex search tasks where users struggle to formulate queries.
\end{abstract}
\section{Introduction}
Being able to retrieve documents in multiple languages is becoming critical as the Internet increasingly provides access to information across a wide range of languages and domains. However, creating effective search queries for cross-language and multi-language retrieval can be a daunting task for users. First, users may be unfamiliar with the language of the documents with the information they need, or may even be unaware of this information, making it hard to craft effective queries. Second, most people are not familiar with the vocabulary and jargon used in other areas or fields, which can hinder their ability to formulate good search queries. 
Consider a scenario where a user is tasked with identifying documents pertinent to legal disputes. They may lack familiarity with the specialized terminology, yet possess examples of specific documents in question.

``Query-by-example'' (QBE) is one solution to such a challenge. It allows users to explore document collections by specifying an example document (rather than an explicit query) of what they are searching for. Although considerable advancements have been made in the domain of query-by-example, in recently using neural IR techniques~\cite{qbe1,qbe2,qbe3,zhang2012query}, there is a lack of effective and easily configurable search interface tools for exploring and annotating query-by-example experiments, especially in the interactive setting. 
\begin{figure*}
  \includegraphics[width=\textwidth]{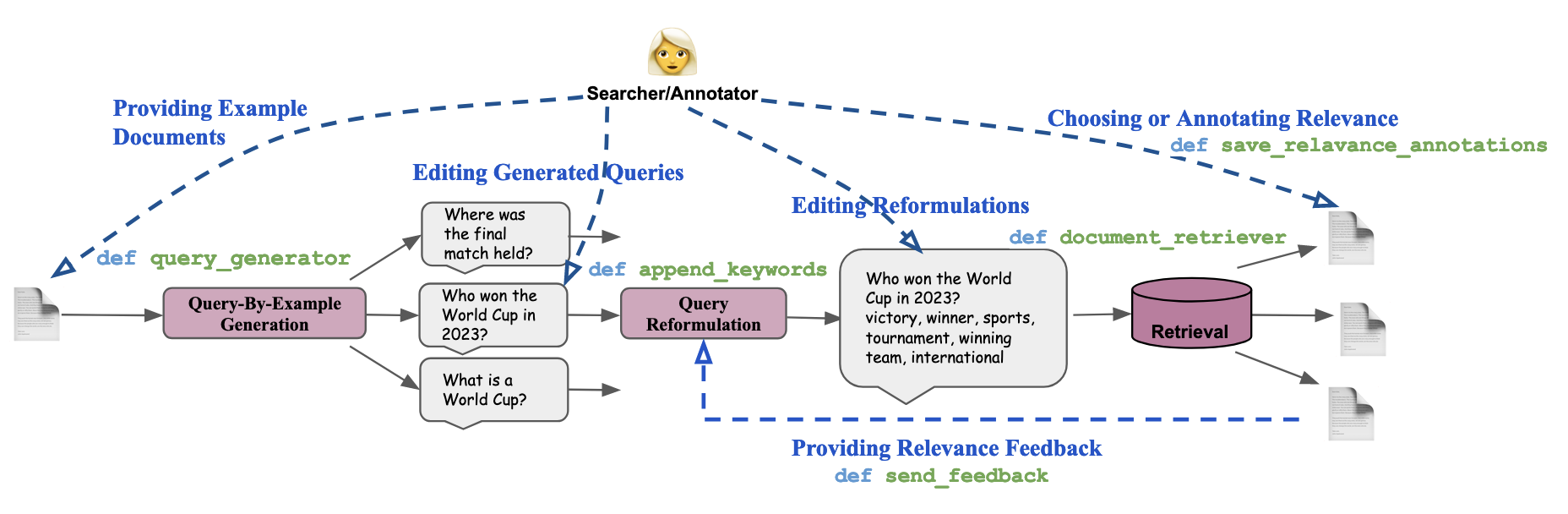}
  \caption{\QueryTuner{}'s process shown end to end along with the internal Python functions. Internal Python helper functions are shown in green, and annotator actions are shown in blue.}
  \label{fig:teaser1}
\end{figure*}

While a good QBE interface could be valuable, a tool that can facilitate generating queries with a Human-In-The-Loop (HITL) setting can result in an even more effective search. Extensive prior work has shown that automatically generated queries can be improved with the searcher's inputs~\cite{jiang2014learning}, via providing in-domain and world-knowledge~\cite{cho-etal-2022-query,mackie2023generative}, as well as from relevance or pseudo-relevance feedback~\cite{abdul2004umass, li-etal-2018-nprf,zheng-etal-2020-bert, wang2023colbert} through search results and this process can iterate several times till the searcher is satisfied with the presented search results. Researchers studying searcher behavior and gathering associated annotated data may need iterating over different query generators, multiple prompting and training strategies, several retrieval pipelines, recording various ways to incorporate implicit feedback, and a rigorous number of hyperparameters. For instance, a search engine analyst might be interested in how searchers edit queries.

To investigate these much-needed capabilities, we propose a configurable, interactive search interface tool, \textbf{\QueryTuner{}}, which supports automatic and interactive query generation and reformulation for mono-lingual and multi-lingual interactive search. \QueryTuner{}  can both facilitate and record the query generation process and interactions of a searcher -- from query formulation in a pure QBE setting to tracking the resulting query's impact on the retrieved search results and the user's exploration. To our knowledge, \QueryTuner{} is the first such interactive query exploration interface to be shared with the research community. 

Specifically, \QueryTuner{} demonstrates the following novel capabilities: 
\begin{itemize}
    \item A simple document search functionality with natural support for multi-lingual and cross-lingual search, making it easy for searchers to navigate and analyze search results.
    \item Support for both automated query generation and reformulation, and human-in-the-loop capabilities such as propagating the human input back to query reformulation, allowing searchers to collaborate with automated query generation models.
    \item Provisioning for rapid prototyping across multiple retrieval experiments and datasets, via PyTerrier retrieval pipelines~\cite{pyterrier} with integrated generative LLM models from HuggingFace~\cite{huggingface}.
    \item Extensive instrumentation support for query generation and reformulation experiments, including the ability to record query edits, reformulations, all user interactions, and relevance judgments making it useful for collecting and annotating end-to-end datasets.
\end{itemize}

In summary, we believe \QueryTuner{} could provide valuable tools for i) performing qualitative analysis over information retrieval experiments and datasets, ii) investigating interactive retrieval feedback and performing Human-In-The-Loop (HITL) studies, and iii) gathering user annotations and understanding searcher behavior. Our system was built initially~\cite{dhole2023interactive} for the BETTER search tasks\footnote{\href{https://www.iarpa.gov/research-programs/better}{IARPA Better Research Program}}~\cite{mckinnon-rubino-2022-iarpa,bettercrosslang} and was later generalized and expanded to support end-to-end query generation and reformulation experiments. We share the Python code, a Google Colab notebook as well as the video demonstration here\footnote{\url{https://github.com/emory-irlab/query-explorer}}. 

Next, we provide an overview of related retrieval tools and the importance of query generation in Section~\ref{relatedwork} to place our contribution in context. We then review the different components and capabilities of \QueryTuner{} in Section~\ref{main}.

\section{Related Work}\label{relatedwork}

There have been many ranking and retrieval tools with annotation support released previously~\cite{pyserini,pyterrier,pyboolir, akiki2023spacerini, ng2023simplyretrieve, giachelle2022doctag}, but all of them have focused on the ad-hoc search setting by assuming a readily available query without the need for generation, reformulation or feedback. Spacerini~\cite{akiki2023spacerini} leveraged the Pyserini~\cite{pyserini} toolkit and the Hugging Face library to facilitate the creation and hosting of search systems for ad-hoc search. SimplyRetrieve~\cite{ng2023simplyretrieve}, FastRag~\cite{fastrag} and RaLLe~\cite{ralle} focus on retrieval augmented generation.

Recent advancements in transformer models and their availability via open-source ecosystems like HuggingFace~\cite{huggingface} and LangChain~\cite{Chase2022LangChain} have facilitated the seamless integration of multiple models~\cite{dhole2024kaucus}. However, despite the accessibility of these tools for researchers and annotators, the integration of the query generator pipeline into search engines remains underdeveloped. Furthermore, the expansion of these tools into multilingual search capabilities has been limited. 

Besides, success in few-shot prompting~\cite{srivastava2022imitation, brown2020language,pretrainprompt,gpt3} has led large language models to play a key role in reducing the information burden on users by especially assisting them with writing tasks namely essay writing, summarisation, transcript and dialog generation, etc. This success has also been transferred to tasks related to query generation~\cite{jeong-etal-2021-unsupervised, nogueira2019document}. While large language model applications are prevalent and numerous studies have been conducted for search interfaces~\cite{s1,s2,s3,s4}, there has been little impetus to combine search interfaces with large language model-based query generation. 

\QueryTuner{} distinguishes itself by offering a more comprehensive integration of various search frameworks, including query generators, reformulators, and multilingual models. Unlike previous approaches, our tool addresses query generation by assuming the `query-by-example' setting, which operates without an explicit query. The query generator component overcomes this challenge by generating a suggested query and refining it through iterative human interaction and feedback.

We now briefly describe the different components of \QueryTuner{}. 

\section{\QueryTuner{}}\label{main}
The \QueryTuner{} Interface is made up of 2 tabs -- The Query Generation tab and the Settings tab. Both of them are described below. The Query Generation tab is displayed to end users or \textbf{searchers} and annotators and the Settings tab is reserved for \textbf{researchers}\footnote{We use the term searchers and researchers to differentiate between the higher level goals of the two tabs but both could encompass analysts/annotators/testers, etc.} looking to gather data for query generation and IR studies by allowing them to investigate different settings. The complete interface is built using HuggingFace's Gradio platform. Gradio~\cite{abid2019gradio} is an open-source Python package to quickly create easy-to-use, configurable UI components and has been popularly used for machine learning models.

\subsection{Searcher's Tab: Query Generation}
This Query Generation tab serves as a simple interface for searchers and annotators which permits end-to-end \textbf{query generation (QG)} -- i) Ad-hoc QG: users can write search queries by themselves ii) Query-by-Example: users can generate queries through prompting a HuggingFace~\cite{huggingface} model and select appropriate ones, \textbf{query reformulation} -- through a HuggingFace model to generate useful keywords, and \textbf{document or passage retrieval} -- through a PyTerrier~\cite{pyterrier} retrieval pipeline over multiple retrieval datasets supported through IRDatasets~\cite{irds}. We display the top-k relevant documents and their source language translations if applicable.

Each of the generated queries can be used by itself or in combination to retrieve documents. Users can further edit the queries, as well as receive assistance from the output of a query reformulator. Users can further interact with the retrieved documents or passages, and provide relevance annotations for each of the documents.

We now describe the default models provided for each of the above settings for the demonstration. Each of these can be easily substituted with the researcher's choices by minor modifications to the configuration settings or code. 

\begin{table*}[htbp]
\centering
\resizebox{2\columnwidth}{!}{
    \centering
    \begin{tabular}{l|l|l}
        \textbf{Functions} & \textbf{Actions (Searchers and Annotators)} & \textbf{Configurable Settings (Researchers)} \\ \hline
        \textbf{ } & Provide Example Documents & Choice of Domain, Example Documents\\ 
        \textbf{query\_generator} & Edit Generated Queries & 0-shot/Few-shot QG, Prompt, Exemplars, HF model \\ 
        \textbf{append\_keywords} & Edit Reformulations to Create Better Query & 0-shot/Few-shot QG, Prompt, Exemplars, HF model \\ 
        \textbf{document\_retriever} & Annotate Relevance of Document to Query & Retrieval PyTerrier Pipeline, Index, Documents, \\ 
        \textbf{send\_feedback} & Select Documents for Providing Relevance Feedback & 0-shot/Few-shot QG, Prompt, Exemplars, Number of Documents \\ 
    \end{tabular}}
    \caption{The different functions (on the left) that searchers and annotators can take assistance from while performing the actions (in the center). Each of them can be configured through the Settings tab along various parameters (shown on the right) by researchers.}
    \label{tab:actions}
\end{table*}

\subsubsection{Query-By-Example Generation (QBE)}
We use the~\texttt{flan-t5-xxl} model~\cite{chung2022scaling}, which is the instruction tuned~\cite{i2} version of the text-to-text transformer T5~\cite{raffel2020exploring}. It has been fine-tuned on a large number of tasks making it convenient~\cite{aribandi2022ext} for learning new tasks. The default version shown to the user is a few-shot wrapper over~\texttt{flan-t5-xxl} -- that takes in the user's example document or passage and prepends an instruction~\texttt{Generate a query given the following document} along with 3 document-query pairs from MSMarco as exemplars to it.

\subsubsection{Query Reformulation (QR)}
We use a zero-shot approach to generate keywords for the given query.~\texttt{flan-t5-xxl} is passed the instruction~\texttt{Improve the search effectiveness by suggesting expansion terms for the query}~\cite{zeroshot} along with the original query as input. Zero-shot query reformulation~\cite{genqrensemble, 10.1145/3578337.3605143, zeroshot} has been recently popular to expand queries to increase their retrieval effectiveness through zero-shot prompting of large language models. A user-facing interface can provide opportunities to mitigate bad reformulations~\cite{whenfail} (Refer Appendix Listing~\ref{lst:qr}).

\subsubsection{Retrieving Documents}
For retrieval, we employ PyTerrier retrievers as default. The architecture of PyTerrier is inherently designed to support operations over retrievers and rerankers to build end-to-end retrieval pipelines and has been a popular choice of retrieval engine among information retrieval researchers. In \QueryTuner{}, researchers can add their own custom PyTerrier pipelines too in the below dictionary (Refer Appendix Listing~\ref{lst:retr}). This would also be reflected in the dropdown in the Settings tab.

\subsubsection{Incorporating Relevance Feedback (RF)}
We provide searchers the ability to improve the current query by utilizing a retrieved document of their choice. We use a zero-shot approach to incorporate the user-selected documents in the style of~\citet{zeroshot, genqrensemble}. The user-selected documents and the query are prompted to regenerate keywords through an instruction~\texttt{Based on the given context information $C$, generate keywords for the following query} where $C$ is a user-selected document (Refer Appendix Listing~\ref{lst:feedback}).

\begin{figure*}
\centering
\fbox{
  \includegraphics[width=0.97\textwidth]{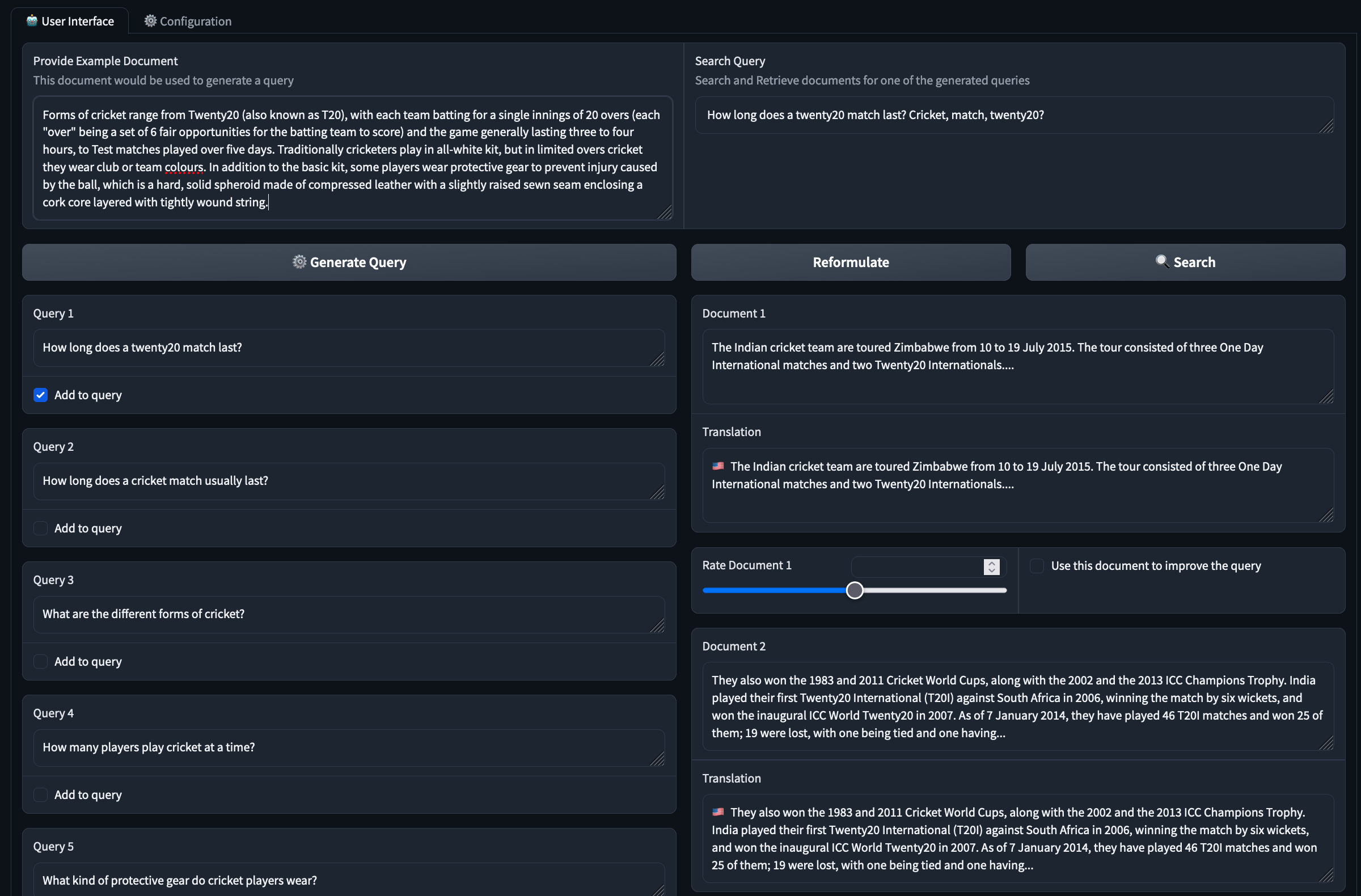}}
  \caption{The User Interface tab: The user provides an example document related to cricket, uses the query generator to generate multiple queries, selects one of them, and uses the reformulator to further improve the query. In this case, the reformulator has suggested a useful term ``cricket'' to increase the retrieval effectiveness of the initial query. }
  \label{fig:searchertab}
\end{figure*}

\subsection{PyTerrier and HuggingFace support}
In an effort to expedite the process of prototyping diverse experiments for IR researchers, \QueryTuner{} incorporates PyTerrier~\cite{pyterrier} support. This integration enables the utilization of retrieval pipelines created through PyTerrier within the \QueryTuner{}  interface, enhancing its functionality for both annotation and qualitative analysis across different retrieval and reranking algorithms. During the demonstration, we showcase the interface's capability to display search results using the BM25 pipeline, highlighting the flexibility to substitute this with other custom pipelines as needed. This feature essentially adds a layer of qualitative analysis to the PyTerrier retrieval pipelines. Furthermore, to broaden the utility of PyTerrier in handling IR datasets, \QueryTuner{}  has been designed to facilitate the indexing of documents from these datasets, thereby enabling qualitative experiments across multiple benchmarks and datasets.

Recognizing the widespread adoption of HuggingFace's~\cite{huggingface} models within the research community, \QueryTuner{} leverages these models for query generation, reformulation, and incorporating feedback. This allows for the comprehensive evaluation of search functionalities across a diverse range of large language models.

While we designed with PyTerrier and HuggingFace ecosystems in mind due to their popularity, datasets and models using other packages can also exploit the \QueryTuner{} interface over their systems.

\subsection{Relevance Annotations}
We allow each document to be annotated for relevance to help researchers gather relevance annotations through a slider component. Annotations are immediately saved in a separate~\texttt{JSON} file.

\subsection{Researcher's Tab: Settings}
The Settings tab is designed for researchers (or specialists) who intend to gather data or study the performance of the interaction by allowing them to vary the various components in the query generation pipeline -- like the choice of retriever, the dataset to retrieve from, or the instruction and few-shot examples for the query generator and reformulator. The various dimensions along which the researcher can vary the settings from the interface and the corresponding searcher's actions are described in Table~\ref{tab:actions}. 

\begin{figure*}
\fbox{
  \includegraphics[width=\textwidth]{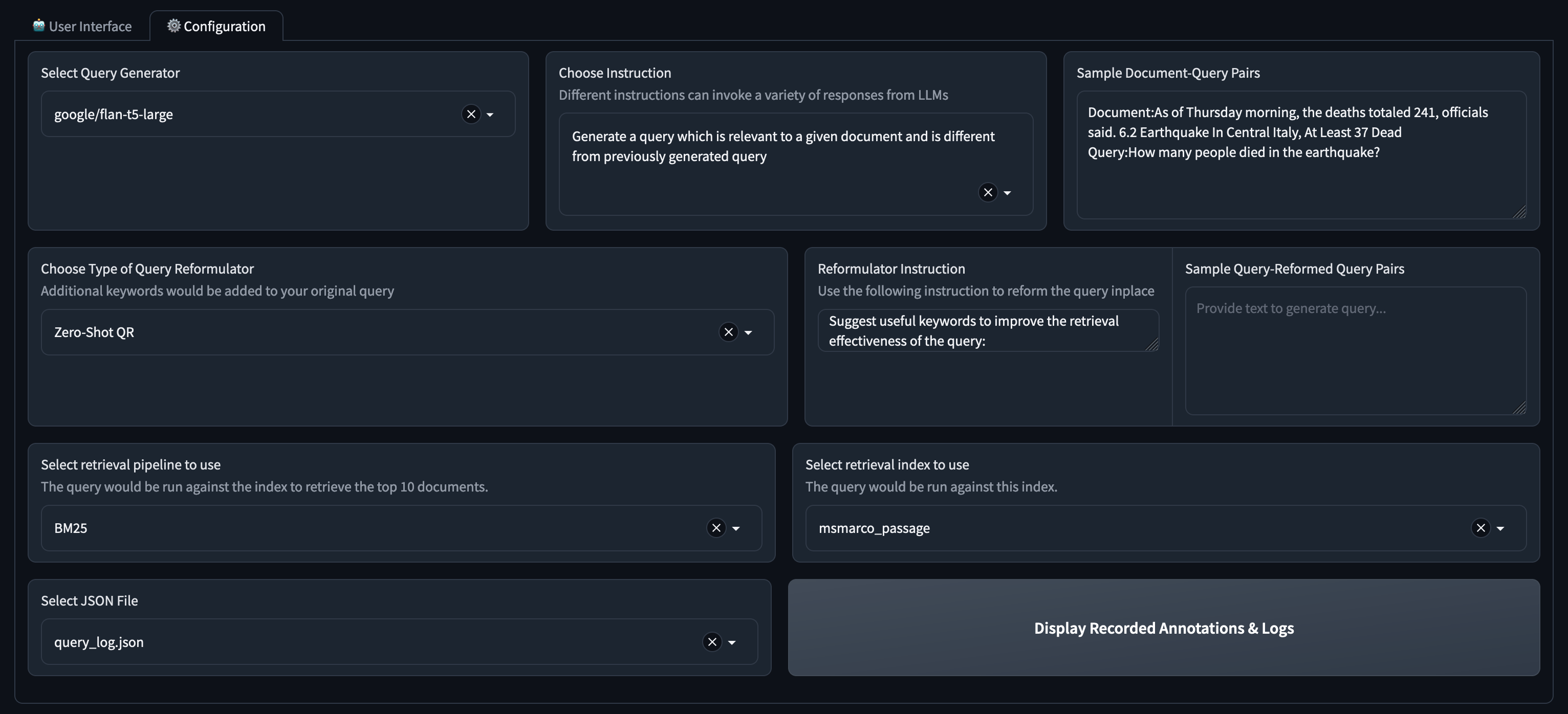}}
  \caption{The Settings Tab where researchers or specialists can experiment with different model settings and parameters and visualize and monitor continuously updated interaction data.}
  \label{fig:settingstab}
\end{figure*}

\subsubsection{Interaction logging}
The researcher can look at the continuously recorded annotations consisting of -- generated queries, post-reformulation queries, query-document relevance annotations, and feedback information -- all with metadata of session and timestamps. These can be viewed directly in the tab as well as be utilized for subsequent analysis.

\QueryTuner{} by default stores the recordings in three \texttt{JSON} formatted files:
\begin{itemize}
    \item \textbf{Query Logs}: where different versions of the queries along with the source of change (whether through a model or the user or through a reformulator, etc.) and additional metadata like timestamps and user session information are stored.
    \item \textbf{Predicted Search Results}: where the user's search queries and corresponding retrieved documents are stores
    \item \textbf{Document-Query Relevance Annotations}: where an annotator's document-query annotated pairs are stored
\end{itemize}

Documenting detailed annotations such as changes in queries and the evolution of queries over time can provide several benefits for researchers. This includes the ability to detect users who may not be attentive or who might be using automated bots. Additionally, observing the patterns and behaviors of users during their search activities can offer valuable insights. Furthermore, assessing the effort involved in formulating queries and perusing documents, as indicated by the time spent on these tasks, can also be advantageous for research purposes.

\section{Conclusion}\label{conclusion}
Interactive query generation and reformulation is of significant interest for many search and exploration tasks like (document-query) pairs augmentation~\cite{qbe3,robust2}, document expansion~\cite{nogueira2019document, gospodinov2023doc2query} and keyword expansion~\cite{genqrensemble, qrsurvey, promptingqueryexpa, zeroshot}. \QueryTuner{} acts as a resource to permit qualitative evaluation of query generation and retrieval in conjunction. Such a combined interface is crucial as it permits immediate retrieval feedback from the user to be incorporated into the search process. 

This paper demonstrates the novel capabilities of \QueryTuner{} to assist researchers with investigating the construction, feedback, and evaluation of the interactive query generation process. Furthermore, \QueryTuner{} provides extensive fine-grained instrumentation to record the end-to-end generation process from query formulation to retrieval feedback to enable the construction of search interaction and feedback datasets. Researchers can also quickly perform qualitative analysis by loading up \QueryTuner{}'s lightweight search interface through Colab and gather data quickly.~\QueryTuner{}'s interface also provides an avenue to perform Human-In-The-Loop (HITL) studies. Apart from qualitative studies, we believe \QueryTuner{} could be effective in performing more sophisticated information retrieval experiments as well as serve as a tool to incorporate retrieval feedback and conduct Human-In-The-Loop studies.

\section{Ethical Statement}
\QueryTuner{} serves as a comprehensive tool, capturing the entire pipeline for purposes of analysis, annotation, and logging. The components within~\QueryTuner{}, including query generators, reformulators, and PRF, can be replaced with even larger LM alternatives. However, these substitutes might lead to the generation of biased or toxic keywords and reformulations. Therefore, it is crucial to consider~\QueryTuner{} within the broader sociotechnical framework~\cite{dhole-2023-large} implement appropriate filters, and conduct thorough testing before any deployment.

\section{Acknowledgements}
This work was supported in part by IARPA BETTER (\#2019-19051600005). The views and conclusions contained in this work are those of the authors and should not be interpreted as necessarily representing the official policies, either expressed or implied, or endorsements of ODNI, IARPA, or the U.S. Government. The U.S. Government is authorized to reproduce and distribute reprints for governmental purposes notwithstanding any copyright annotation therein.

\bibliography{custom}

\appendix
\section{Appendix}\label{sec:appendix}
\begin{lstlisting}[caption={Standalone Query Reformulation using zero-shot prompting}, label=lst:qr]
def append_keywords(session, query, reform_method, reform_instruction, hf_model, file_name='query_reformulations'):

  rf_queries = query_generator(f'Generate keywords for the query : ','', query, hf_model, session)
  
  reformed_query = query + ' ' + rf_queries[1]
  
  on_query_change(reformed_query, file_name, session, previous_query=query) # Record event
  
  return ref_query
\end{lstlisting}

\begin{lstlisting}[caption={Triggering Retrieval: Researchers can extend the dictionary using their custom pipelines.}, label=lst:retr]

retrieval_algos_dict = {'BM25': bm25, 'TF_IDF': tfidf}

def retrieve_for_ui(query_text, pipeline=bm25):

  # User Query used to retrieve through a PyTerrier Pipeline
  searchresults = (pipeline%10).search(cleanup(query_text))

  # Document text for display
  searchresults['eng-text'] = searchresults['docno'].apply(get_doc_text)

  # (Optional) Translation for cross/multi-lingual
  searchresults['target-text'] = translate(searchresults['eng-text'], 'eng', 'eng')
  
  results = [row.to_dict() for index, row in searchresults1.iterrows()]
  
  return results
\end{lstlisting}

\begin{lstlisting}[caption={Query Reformulation With Relevance Feedback using Zero-shot prompting}, label=lst:feedback]
def send_feedback(query, document, hf_model, session, file_name='feedback_query_reformulations'):
  
  rf_queries = query_generator1(f'Based on the given context ```{document}```, generate keywords for the query : ', query, hf_model, session)
  
  ref_query = query + " " + rf_queries[1]
  
  on_query_change(ref_query, file_name, session, previous_query=query) # Record event
  
  return ref_query
\end{lstlisting}

\end{document}